\documentclass[11pt,a4paper]{article}

\usepackage{jheppub}

\usepackage{subfigure}
\usepackage{latexsym,amsmath,amssymb}
\allowdisplaybreaks

\title{Two critical phenomena in the exactly soluble quantized Schwarzschild black hole} 

\author[a]{Edwin J. Son,}
\author[b,c]{and Wontae Kim}

\affiliation[a]{Division of Computational Sciences in Mathematics,
National Institute for Mathematical Sciences, Daejeon 305-390, Republic of Korea}
\affiliation[b]{Department of Physics, Sogang University, Seoul, 121-742, Republic of Korea}
\affiliation[c]{Center for Quantum Spacetime, Sogang University, Seoul 121-742, Republic of Korea}

\emailAdd{eddy@nims.re.kr}
\emailAdd{wtkim@sogang.ac.kr}

\abstract{%
We study thermodynamic quantities and phase transitions 
of a spherically symmetric Schwarzschild black hole by taking into account the back reaction through the conformal anomaly of matter fields,
and show that there exists an additional phase transition
to the conventional Hawking-Page phase transition. 
The small black hole is more probable than the hot flat space above a second critical 
temperature, while it is less 
probable than the hot flat space in the classical Schwarzschild 
black hole. However, the unstable small black hole eventually 
should decay into the stable large black hole 
because the conformal anomaly does not change its thermodynamic stability. 
}

\keywords{Black Hole, Thermodynamics}

\arxivnumber{1212.2307}

\begin{document}

\maketitle

\newcommand{\lp}{\ell_P}

\section{Introduction}
\label{sec:intro}
Since the area law of the black hole entropy  has been discovered 
 by Bekenstein and Hawking~\cite{Bekenstein:1973ur,Bekenstein:1974ax,Hawking:1974sw}, 
there has been much attention to thermodynamic quantities and thermodynamic phase transitions 
on various black holes;
Schwarzschild-AdS type black holes~\cite{Hawking:1982dh,Brown:1994gs,Cai:2001dz,Cvetic:2001bk,Nicolini:2011dp,Banerjee:2011au}, 
black branes~\cite{Gubser:1996de,Cai:2007vv,Cadoni:2012uf}, and
lower dimensional black holes~\cite{Nappi:1992as,Lemos:1996bq,Zaslavskii:2003cr,
Grumiller:2007ju,Quevedo:2008ry,Clement:2009gq,Birmingham:2010mj} together with the
statistical relation of the entropy~\cite{'tHooft:1984re,Frolov:1994zi,Carlip:1998wz,Kim:2008bf}.
In particular, as for the conventional thermodynamic properties of the Schwarzschild black hole, 
a stable large black hole and an unstable small black hole appear above the minimum of local temperature.
The unstable small black hole can decay into either the large black hole or pure thermal 
radiation~\cite{Hawking:1982dh,York:1986it,Whiting:1988qr,Gross:1982cv}.
In the large black hole, the Hawking-Page (HP) phase transition appears
since the free energy of the black hole is lower than the free energy of the hot flat space.
Since all these results have been based on the classical metric without the back reaction
of the geometry, it seems to be natural to ask what happens if we take into account  
the quantum back reaction of the metric in order to investigate the thermodynamics and
the phase transition. 

However, it is not easy to study the back-reacted 
thermodynamic quantities analytically even in the semi-classical level. 
First of all, the renormalizable quantum corrections are 
in general impossible in the Einstein-Hilbert action.
Even though one can restrict just one-loop corrections, 
it is intractable to get the exact solution from the 
equation of motion~\cite{York:1984wp}.
Recently, a class of exact analytic and static solution has been derived 
from the semi-classical Einstein equation with the conformal anomaly,
and the modification of the entropy has been discussed~\cite{Cai:2009ua}.  
So, it would be interesting to study whether the classical phase transition can be modified or not via the exactly soluble model presented in Ref.~\cite{Cai:2009ua}. Thus, we would like to calculate the quantum corrected thermodynamic quantities 
and study phase transitions, especially paying attention to the small black hole where the
quantum effects are significant.
As expected, the HP phase transition from the hot flat space into the large black hole 
occurs above the first critical temperature.
Moreover, it turns out that 
there exists new kind of phase transition from the hot flat space to the small
black hole at the second critical temperature
which is actually impossible in the classical metric. 

The paper is organized as follows.
The exactly soluble metric with the quantum back reaction~\cite{Cai:2009ua} will be introduced in section~\ref{sec:solution}, 
and then we shall consider the special configuration of the metric from
general solutions 
by choosing appropriate integration constants. The advantage of this choice is that the flat metric from the   
quantum-corrected one can be regarded as the vacuum. 
In section~\ref{sec:thermo}, quantum corrected thermodynamic quantities will be calculated. In particular, the entropy receives a logarithmic correction~\cite{Cai:2009ua}.
Introducing the finite cavity as an isothermal surface~\cite{York:1986it},
the thermodynamic energy and the heat capacity will be calculated
based on the modified entropy which consists of
the ordinary area term and the logarithmic term.  
In section~\ref{sec:freeenergy}, the HP phase transition will be studied,
so that in the large black hole the first phase transition occurs
at the first critical temperature. Moreover, it can be shown that
the second phase transition from the hot flat space 
to the small black hole appears at the second critical temperature which is larger than
the first critical temperature. We will point out some differences between 
these two phase transitions. Finally, summary and discussion are 
given in section~\ref{sec:discuss}.

\section{A black hole solution with conformal anomaly}
\label{sec:solution}
Let us start with an exact spherically symmetric black hole solution 
in the semi-classical Einstein equation with the conformal anomaly~\cite{Cai:2009ua}.
The back-reaction of quantum fields to the space-time geometry can be
included in the curved space-time~\cite{Birrell:1982ix};
\begin{equation}
\label{eom}
  G_{\mu\nu}= 8\pi G_N \langle T_{\mu\nu} \rangle,
\end{equation}
where $G_{\mu\nu}=R_{\mu\nu} - \frac12 g_{\mu\nu} R $ and 
$\langle T_{\mu\nu} \rangle$ is the effective energy-momentum tensor.
In four dimensions, a trace anomaly from the energy-momentum tensor can be written as~\cite{Deser:1993yx,Duff:1993wm}
\begin{equation}
\label{anomaly}
g^{\mu\nu} \langle T_{\mu\nu} \rangle = -\frac{\alpha}{8 \pi} E_{(4)} + \tilde{\alpha}
 C_{\alpha\beta\kappa\lambda} C^{\alpha\beta\kappa\lambda}, \\
\end{equation}
where the coefficients $\alpha$ and $\tilde{\alpha}$ depend on matter contents,
$E_{(4)} = R_{\mu\nu\kappa\lambda} R^{\mu\nu\kappa\lambda} - 4 R_{\mu\nu} R^{\mu\nu} + R^2$
is the Gauss-Bonnet term, and $C_{\alpha\beta\kappa\lambda}$ is the Weyl tensor.

By assuming $\tilde{\alpha}=0$ for the exact solubility with
the relations of $\langle T^t_{~t} \rangle = \langle T^r_{~r} \rangle$ in the whole space-time \cite{Cai:2009ua},  one can find the class of exact solutions for a static and spherically symmetric space-time of
 $ds^2 = -f(r) dt^2 +dr^2/f(r) + r^2 d\Omega_2^2$;
\begin{equation}
\label{sol:f}
f(r) = 1 - \frac{r^2}{4\alpha} \left[ 1 - \sqrt{1 - \frac{16\alpha M}{r^3} + \frac{8\alpha b}{r^4}} \right],
\end{equation}
where $M$ and $b$ are integration constants, $\alpha>0$, and we set the Newton constant $G_N = 1$. 
There are largely two classes of solution depending on $b$.
For $b >0$,
the asymptotic behavior of the metric~\eqref{sol:f} is given by
\begin{equation}
\label{asymp:sol}
f \approx 1 - \frac{2M}{r} + \frac{Q^2}{r^2} + O(r^{-4})
\end{equation}
by setting $b=Q^2$, so that one can imagine that it approaches the Reissner-Nordstr\"om black hole
with the mass $M$ and the electric charge $Q$ asymptotically.
The properties of the metric and the thermodynamics have been intensively studied in Ref.~\cite{Cai:2009ua}.
For $b <0$,  the asymptotic behavior of the metric is written as
\begin{equation}
\label{f:asymp}
f \approx 1 - \frac{2M}{r} - \frac{a^2}{2r^2} + O(r^{-4})
\end{equation}
by choosing $b=-a^2/2<0$. 
Assuming $a^2$ is the renormalized Newton constant, it turns out that 
it is equivalent to the asymptotic form of the metric obtained from the spherically symmetric 
quantization of the Schwarzschild metric~\cite{Kazakov:1993ha},
where the thermodynamic behavior for the metric was recently investigated in Ref.~\cite{Kim:2012cm}.
On the other hand, in the (canonical) fixed-charge ensemble for a charged black hole,
the HP type phase transition between a black hole and a `hot empty space' does not exist 
but it is possible only for \emph{uncharged} black hole, since there is no
`hot empty charged space'~\cite{Carlip:2003ne}.
So, in the present black hole, we  will take
a charge-free canonical ensemble by taking $b=0$
in order to find HP phase transitions between the
quantum back-reacted black hole
and the hot flat space. 
Therefore, the vacuum of the metric can be easily defined not only by setting $M=0$  but also at the asymptotic infinity,
which is nothing but the
Minkowski space-time.   

Now, the given metric~\eqref{sol:f} has two roots of $r_\pm = M  \pm \sqrt{M^2+2\alpha}$ satisfying $f(r_\pm)=0$;
however, $r_-$ becomes negative for $\alpha>0$, so that the 
single horizon is obtained as
\begin{equation}
\label{horizon}
r_+ = M  +\sqrt{M^2+2\alpha}.
\end{equation}
Note that the horizon $r_+$ is bounded from below,
\begin{equation}
\label{cond:hor}
r_+ \ge r_\text{min} \equiv 2 \sqrt{\alpha},
\end{equation}
from the reality condition of the metric function~\eqref{sol:f}.
The lower bound of the mass parameter corresponding to the minimum horizon
$r_\text{min}$ is given by $M_\text{min} = \sqrt{\alpha}/2$.

\begin{figure}[pt]
  \begin{center}
  \includegraphics[width=.7\textwidth]{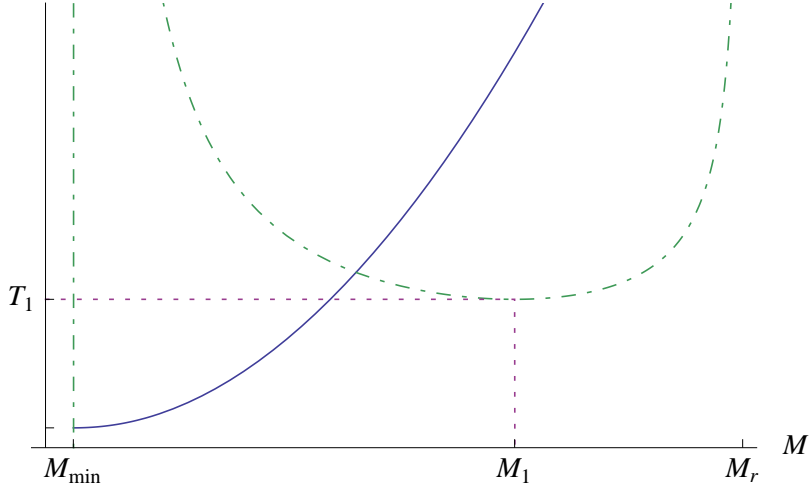}
  \end{center}
  \caption{The entropy (solid curve) and the local temperature (dot-dashed curve) are plotted for $r=2$.
    The entropy is monotonically increasing function while the local temperature has a minimum at $M_1$.} 
   \label{fig:SnT}
\end{figure}

\section{Thermodynamic quantities}
\label{sec:thermo}
We now calculate thermodynamic quantities in a cavity in order to study 
the thermodynamic stability and phase transitions of the quantum corrected black hole 
described by the metric by setting $b=0$ in Eq.~\eqref{sol:f}.
First of all, the entropy of the black hole is given by the thermodynamic first law~\cite{Cai:2009ua}
\begin{equation}
\label{S}
S = \frac{A}{4} - 4 \pi \alpha \ln \frac{A}{A_0},
\end{equation}
where $A = 4 \pi r_+^2$ is the area of the horizon and $A_0$ reflects the logarithmic ambiguity
which can be chosen as $4\lp^2$ for convenience.
The area law of the entropy can be recovered by setting $\alpha = 0$.
Since the horizon radius $r_+$ is bounded from below, so is the area $A$, and the entropy remains finite as
 $M$ goes to the minimum $M_\text{min}$.
The profile of the entropy~\eqref{S} is presented in Fig.~\ref{fig:SnT}.

Next, the finite cavity of the isothermal surface is employed to investigate thermodynamic behaviors
so that the present black hole is located at the center 
of a spherical cavity of radius $r$. 
The corresponding local temperature for the observer on the cavity 
is written as~\cite{Tolman:1930zza} 
\begin{equation}
\label{T}
T = \frac{\sqrt{M^2+2\alpha}}{2\pi [ (M+\sqrt{M^2+2\alpha})^2-4\alpha] \sqrt{f(r)}},
\end{equation}
where it recovers the Hawking temperature for the infinite cavity.
The local temperature~\eqref{T} is plotted in Fig.~\ref{fig:SnT}.
Note that $T$ diverges when $M$ approaches either $M_\text{min}$ or  
$M_r$.
The minimum of the local temperature $T_1$ yields a critical mass of $M=M_1$,
where $M_1$ can be numerically obtained if needed.

Integrating the thermodynamic first law of $dE = T dS$, 
one can obtain the local energy, 
\begin{equation}
\label{E}
\begin{aligned}[b]
E &= E_0 + \int T dS \\
&\begin{aligned} =\ r \left[ 1 - \left(1 - \frac{4\alpha}{r^2}\right) \sqrt{f(r)} \right]
 - \frac{4\alpha}{3r} f^{3/2}(r),
 \end{aligned}
\end{aligned}
\end{equation}
where we chose $E_0 = r$ to make $E = M$ as $r\to\infty$.
The energy~\eqref{E} increases monotonically as $M$ grows,
which can be easily seen from the fact that the metric function $f(r)$ is a monotonically decreasing function.
Thus, the behavior of the energy~\eqref{E} looks more or less similar to the classical black hole case
of $E=r -r \sqrt{1 -2M/r}$ for the large black hole; however, 
the quantum effects becomes significant for the small black hole.
For instance, for the extreme case of $M \to M_{\text{min}}$, the
energy becomes $E_{\text{min}} = r - \sqrt{f} \left[ r - (4\alpha/r) ( 1 - f/3) \right]$ 
which is still positive since the metric function is in the range of $0<f<1$ outside the horizon $r>r_+$.

Now, one might wonder how to derive the thermodynamic energy \eqref{E} from the fundamental
relation and eventually verify the thermodynamic first law. One of the best way to get
thermodynamic quantities is to use Euclidean action formalism~\cite{Hawking:1982dh,Gibbons:1976ue,Gibbons:1978ac} 
or the Hamiltonian formulation~\cite{Brown:1994gs}; however, this is not
the case since we are not well aware of the action and the Hamiltonian  as seen from the complication of
Eq.~\eqref{anomaly}. Instead, we would like to derive the thermodynamic first law directly from the
staring equation of motion~\eqref{eom}.
The relevant tensor component for the static spherically symmetric solution~\eqref{sol:f}
is only radial part,
\begin{equation}
\label{radial equation}
G^r_{~r}-8\pi <T^r_{~r}>=0,
\end{equation}
where the Einstein tensor and the anomaly term can be written as
\begin{align}
G^r_{~r}(r)&=-\frac{1}{r^2}(1-f-rf'),\\
<T^r_{~r}>&=\frac{\alpha}{4\pi r^4}(1-f)(1-f+2rf'),
\end{align}
respectively. The prime means the derivative with respect to
the radial coordinate.
Considering this equation at the event horizon defined by $r = r_{+}$
in Eq. \eqref{horizon}, one can obtain equation of motion defined at the horizon.
Next, we multiply~\eqref{radial equation} by $(-r_+^2 dr_{+} /2)$ based on the method to relate  
thermodynamic relations with the Einstein equation of motion~\cite{Padmanabhan:2012gx},
which gives 
\begin{equation}
\label{crude first law} 
d(\frac{r_{+}}{2}-\frac{\alpha}{r_{+}})-\frac{f'(r_{+})}{4\pi} d(\frac{A}{4}-4\pi\alpha \ln \frac{A}{4} + S_0)=0,
\end{equation}
where $A=4\pi r_{+}^2$. Note that there are $\alpha$-correction at each term
due to the anomaly.
From Eq.~\eqref{horizon}, we can express $M$ in terms of the horizon as $M=r_{+}/2-\alpha/r_{+}$,
then Eq.~\eqref{crude first law} becomes
\begin{equation}
dM-T_{\rm H }dS=0
\end{equation}
as long as we identify the Hawking temperature $T_{\rm H}=f'(r_{+})/4\pi$ which
is just the surface gravity along with an appropriate normalization factor.
Thus, it is natural to identify the entropy with $S=A/4-4\pi\alpha \ln A/4 + S_0$
which is the same with Eq.~\eqref{S} for $S_0 = 0$.
Moreover, it is obvious that the thermodynamic energy is the ADM mass $M$,
which can be also read from the asymptotic solution~\eqref{asymp:sol}.
This thermodynamic first law can be embedded into thermodynamics with the cavity in order to 
obtain the thermodynamic local energy $E_{\text{loc}}$. 
It can be realized by considering the red-shift factor due to the
cavity effect, that is,  
\begin{equation}
\frac{1}{\sqrt{f(r)}}dM-\frac{T_H}{\sqrt{f(r)}}dS=0.
\end{equation}
Let us define $dE_{\text{loc}}\equiv \frac{1}{\sqrt{f(r)}}dM$, and then we can get
\begin{equation}
dE_{\text{loc}}=TdS.
\end{equation}
By integrating the local energy with respect to $M$,
it can be shown that the present local thermodynamic energy $E_{\text{loc}}$  is 
equivalent to the previous thermodynamic energy~\eqref{E}: $E_{\text{loc}}\equiv E$.

\begin{figure}[pt]
  \begin{center}
  \includegraphics[width=.7\textwidth]{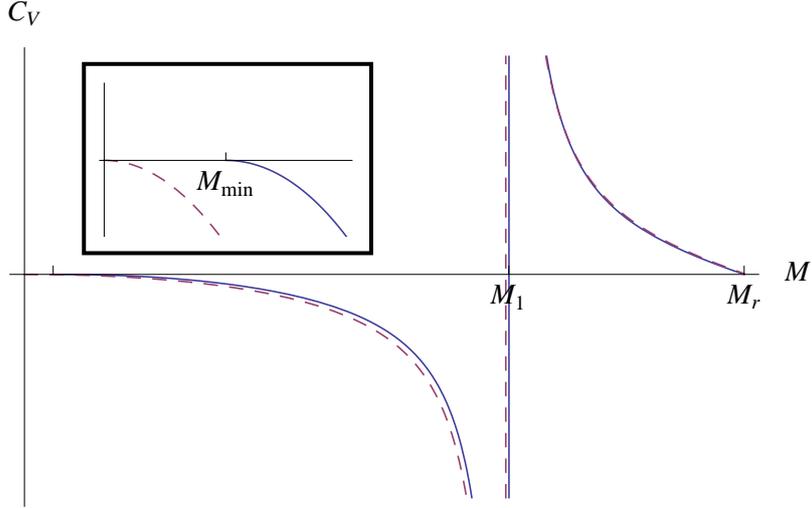}
  \end{center}
  \caption{The heat capacity (solid curve) is plotted for $r=2$, which is 
     is similar to the classical Schwarzschild black hole (dashed curve).
    The specific heat is zero at $M_\text{min}$.} 
  \label{fig:heatcapacity}
\end{figure}

Finally, for the thermodynamic stability of the black hole, 
one can simply calculate the heat capacity as
\begin{equation}
\label{CV}
C_V = \left( \frac{\partial E}{\partial T} \right)_r
= \frac{2\pi \gamma \delta \left( r_+^2 - 4\alpha \right)^2 f}{\left( \gamma^2/r + \delta M f \right) \left( r_+^2 - 4\alpha \right) - 2 \gamma \delta r_+^2 f},
\end{equation}
where $\gamma=\sqrt{M^2 + 2\alpha}$ and $\delta=\sqrt{1- 16\alpha M/r^3}$.
The heat capacity is divergent at the critical mass $M_1$. 
The profile of the heat capacity is presented in Fig.~\ref{fig:heatcapacity}, 
which is similar to the conventional one except the behavior at $M_\text{min}$.

Actually, it was argued that there is a single large black hole in an asymptotically flat
black hole space-time~\cite{Gross:1982cv}. By the way, it has been claimed that a cavity
in such a space-time can play a similar role to the negative vacuum energy of AdS, so that
there also exist two branches in our case: one is the small black hole and the other is
the large black hole. Moreover, there is an advantage to introduce the finite cavity
in the asymptotically flat black hole. The divergent density of states is compensated with
the weight defined in the cavity, so that the partition function is well-defined~\cite{York:1986it}.
As seen from Fig.~\ref{fig:heatcapacity}, the large black hole appears in the region of
the positive heat capacity, while the small black hole appears in the negative heat capacity.
It means that the essential thermodynamic behaviors of the asymptotically flat black hole observed at
the fixed radial distance would be the same with those of the AdS black hole.

\section{Free energy and phase transition}
\label{sec:freeenergy}
\begin{figure}[pt]
  \begin{center}
  \subfigure[{Schwarzschild black hole}]{\includegraphics[width=0.45\textwidth]{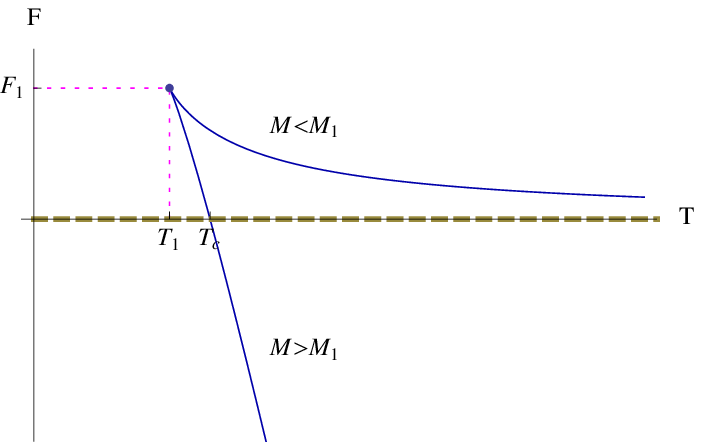}\label{fig:TvF1}}
  \subfigure[{Quantum black hole}]{\includegraphics[width=0.45\textwidth]{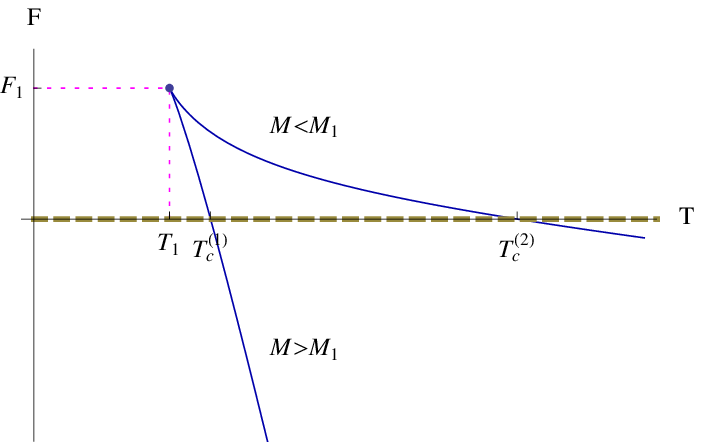}\label{fig:TvF2}}
  \end{center}
  \caption{The on-shell free energies for the unstable 
  small black hole of $M<M_1$ (upper curve) and the stable large 
  black hole of $M>M_1$ (lower curve) are shown for $r=2$. For the case of~\subref{fig:TvF1},
  the on-shell free energy for the small black hole remains positive for the Schwarzschild black hole
  ($\alpha = 0$), while it crosses the horizontal axis at 
  $T_c^{(2)}$ for the case of~\subref{fig:TvF2}.
  The dashed line on the horizontal axis represents the free energy of the hot flat space-time.
 } 
  \label{fig:TvF} 
\end{figure}

Let us define the on-shell free energy of the black hole in order to study the phase transition. 
Using Eqs.~\eqref{S}, \eqref{T} and \eqref{E}, 
the free energy of the black hole is easily obtained 
as $F=E-TS$, 
where its behavior is presented in Fig.~\ref{fig:TvF}.
The free energy of the hot flat space is $F_{\text{flat}}=0$ 
since the vacuum state is Minkowski space-time.
Note that we have fixed $b=0$ to study the phase transition between the 
black hole and the flat space. Actually, 
both metrics satisfy the same asymptotic geometries.

As seen from Fig.~\ref{fig:TvF1} for the ordinary Schwarzschild black hole,
the small and the large black holes are degenerate at $T_1$ corresponding to the critical mass $M_1$.
These two states are higher than the free energy of the hot flat space for $T_1 < T < T_c$, where
$T_c$ is the first critical temperature. 
For $T> T_c$, the free energy of the unstable small black hole of $M<M_1$ is higher than that of the hot 
flat space whereas the free energy of the stable large black hole of $M> M_1$ 
is less than the free energy of 
the hot flat space. Consequently, at the temperature $T>T_c$
 the radiation can collapse to the large black hole through the phase transition 
 and the small black hole can decay into the large black hole thermodynamically. 

Now, as for the case of Fig.~\ref{fig:TvF2}, the non-vanishing $\alpha$ yields
the second phase transition at $T_c^{(2)}$ 
along with the ordinary phase transition at $T_c^{(1)}$.
Thus, if $T>T_c^{(2)}$, the radiation can collapse not only into the large black hole but also
 into the small black hole which is impossible in the classical
thermodynamics of the Schwarzschild black hole.
However, the small black hole is unstable 
as seen from the behavior of the heat capacity, so that it should 
eventually decay into the
large black hole  
because the free energy of the large black hole is still lower than the free energy
of the small black hole. 

As for the classical limit, the solution~\eqref{sol:f} becomes Schwarzschild solution
for the  limit of vanishing $\alpha$, since we set $b=0$ in our analysis. Next, it can be
easily seen that the entropy~\eqref{S} and the temperature~\eqref{T} recover those of
Schwarzschild black hole for this limit: the entropy becomes nothing but the well-known
area law and the  temperature becomes $1/[8 \pi M \sqrt{f(r)}]$ which is the local temperature
of the Schwarzschild black hole. Also, the energy~\eqref{E} becomes $r - r \sqrt{f(r)}$
which is the  local energy of the Schwarzschild black hole. Finally, it follows that
the free energy $F = E - T S$ should become the free energy of the Schwarzschild black hole.
One can see from Fig.~\ref{fig:TvF} that the quantum case~\subref{fig:TvF2} becomes
the case of~\subref{fig:TvF1} for the Schwarzschild black hole as $T_c^{(2)}$ goes to infinity,
which occurs when $\alpha$ tends to zero.

\section{Discussion}
\label{sec:discuss}

We have studied the thermodynamic quantities and the phase transitions 
of the spherically symmetric black hole by taking into account 
the back reaction of the Schwarzschild black hole in terms of the conformal anomaly
of matter fields. So it has been shown that 
there can exist an additional phase transition from the
hot flat space to the small black hole at the second  critical temperature,
which is compared to the ordinary Schwarzschild black hole
where the HP phase transition appears only for the large black hole 
since the free energy of the hot flat space is higher than the free energy of the 
large black hole just in that region. Of course, the small black hole 
eventually decays into the large stable black hole
because this small black hole is unstable.
In fact, the non-triviality of quantum effect comes from 
the $\alpha$-dependent logarithmic correction~\eqref{S} whose form is not unique 
in that it generates the constant entropy contribution of $S_0 = 4\pi\alpha \ln A_0$,
where we have chosen $A_0=4\lp^2$. 
Actually, it has something to do with
the third law of the black hole thermodynamics; however, it has not been well appreciated yet~\cite{Dabholkar:2012pk}.

One can study the off-shell free energy from
the on-shell free energy of the ordinary Schwarzschild black hole in Fig.~\ref{fig:TvF1}.
It shows that the off-shell free energy can be consistently defined along the mass of the black hole since
the sequence of stable ($F=0$ with $M=0)$, unstable ($F>0$  with $M<M_1$), and
stable ($F >0$ or $F<0$ with $M>M_1$) states alternatively appear.
Moreover, the free energy of the small black hole is positive and higher than that of the 
large black hole. There exists a single unstable state between the hot flat space and the large 
black hole state, so that the off-shell free energy is everywhere smooth. 
On the other hand, at a temperature $T>T_c^{(2)}$ on the quantized black hole from Fig.~\ref{fig:TvF2},
all free energy is negative. What it means is that if we want to connect 
the free energy of the hot flat space ($F=0$ with $M=0$) with the free energy of the unstable small black hole 
($F <0$ with $M < M_1$) smoothly in the region of the negative free energy,
we should require a locally stable state
where its free energy is also negative.
Then, we can expect it may be a remnant whose mass is $M_{\text{min}}$;
however, it is somewhat awkward because it is half unstable 
around $M_{\text{min}}$ 
as shown in Fig.~\ref{fig:heatcapacity}.
This intriguing problem could not be addressed in this work because it is non-trivial task to realize the 
off-shell free energy without the explicit action.

\acknowledgments 
W. Kim would like to thank Myungseok Eune, Shailesh Kulkarni, and Jaehoon Jeong for exciting discussions,
and especially Yongwan Kim for valuable comments. 
W. Kim was supported by the National Research Foundation of Korea(NRF)
grant funded by the Korea government (MEST) (2012-0002880), and
by the National Research Foundation of Korea(NRF) grant funded 
by the Korea government(MEST) through the Center for Quantum 
Spacetime(CQUeST) of Sogang University with grant number 2005-0049409.


\begin{thebibliography}{99}

\bibitem{Bekenstein:1973ur}
  J.~D.~Bekenstein,
  \textit{Black holes and entropy,}
  Phys.\ Rev.\ D {\bf 7} (1973) 2333.

\bibitem{Bekenstein:1974ax}
  J.~D.~Bekenstein,
  \textit{Generalized second law of thermodynamics in black hole physics,}
  Phys.\ Rev.\ D {\bf 9} (1974) 3292.

\bibitem{Hawking:1974sw}
  S.~W.~Hawking,
  Commun.\ Math.\ Phys.\ {\bf 43} (1975) 199.

\bibitem{Hawking:1982dh}
  S.~W.~Hawking and D.~N.~Page,
  \textit{Thermodynamics of Black Holes in anti-De Sitter Space,}
  Commun.\ Math.\ Phys.\  {\bf 87} (1983) 577.

\bibitem{Brown:1994gs}
  J.~D.~Brown, J.~Creighton and R.~B.~Mann,
  \textit{Temperature, energy and heat capacity of asymptotically anti-de Sitter black holes,}
  Phys.\ Rev.\ D {\bf 50} (1994) 6394  [gr-qc/9405007].

\bibitem{Cai:2001dz}
  R.~-G.~Cai,
  \textit{Gauss-Bonnet black holes in AdS spaces,}
  Phys.\ Rev.\ D {\bf 65} (2002) 084014  [hep-th/0109133].

\bibitem{Cvetic:2001bk}
  M.~Cvetic, S.~Nojiri and S.~D.~Odintsov,
  \textit{Black hole thermodynamics and negative entropy in de Sitter and anti-de Sitter Einstein-Gauss-Bonnet gravity,}
  Nucl.\ Phys.\ B {\bf 628} (2002) 295  [hep-th/0112045].

\bibitem{Nicolini:2011dp}
  P.~Nicolini and G.~Torrieri,
  \textit{The Hawking-Page crossover in noncommutative anti-deSitter space,}
  JHEP {\bf 1108} (2011) 097
  [arXiv:1105.0188 [gr-qc]].

\bibitem{Banerjee:2011au}
  R.~Banerjee and D.~Roychowdhury,
  \textit{Thermodynamics of phase transition in higher dimensional AdS black holes,}
  JHEP {\bf 1111} (2011) 004
  [arXiv:1109.2433 [gr-qc]].

\bibitem{Gubser:1996de}
  S.~S.~Gubser, I.~R.~Klebanov and A.~W.~Peet,
  \textit{Entropy and temperature of black 3-branes,}
  Phys.\ Rev.\ D {\bf 54} (1996) 3915
  [hep-th/9602135].

\bibitem{Cai:2007vv}
  R.~-G.~Cai, L.~-M.~Cao and Y.~-W.~Sun,
  \textit{Hawking-Page Phase Transition of black Dp-branes and R-charged black holes with an IR Cutoff,}
  JHEP {\bf 0711} (2007) 039  [arXiv:0709.3568 [hep-th]].

\bibitem{Cadoni:2012uf}
  M.~Cadoni and S.~Mignemi,
  \textit{Phase transition and hyperscaling violation for scalar Black Branes,}
  JHEP {\bf 1206} (2012) 056  [arXiv:1205.0412 [hep-th]].

\bibitem{Nappi:1992as}
  C.~R.~Nappi and A.~Pasquinucci,
  \textit{Thermodynamics of two-dimensional black holes,}
  Mod.\ Phys.\ Lett.\ A {\bf 7} (1992) 3337  [gr-qc/9208002].

\bibitem{Lemos:1996bq}
  J.~P.~S.~Lemos,
  \textit{Thermodynamics of the two-dimensional black hole in the Teitelboim-Jackiw theory,}
  Phys.\ Rev.\ D {\bf 54} (1996) 6206  [gr-qc/9608016].

\bibitem{Zaslavskii:2003cr}
  O.~B.~Zaslavskii,
  \textit{Two-dimensional quantum corrected black hole in a finite size cavity,}
  Phys.\ Rev.\ D {\bf 69} (2004) 044008  [hep-th/0310268].

\bibitem{Grumiller:2007ju}
  D.~Grumiller and R.~McNees,
  \textit{Thermodynamics of black holes in two (and higher) dimensions,}
  JHEP {\bf 0704} (2007) 074  [hep-th/0703230].

\bibitem{Quevedo:2008ry}
  H.~Quevedo and A.~Sanchez,
  \textit{Geometric description of BTZ black holes thermodynamics,}
  Phys.\ Rev.\ D {\bf 79} (2009) 024012  [arXiv:0811.2524 [gr-qc]].

\bibitem{Clement:2009gq}
  G.~Clement,
  \textit{Warped AdS(3) black holes in new massive gravity,}
  Class.\ Quant.\ Grav.\  {\bf 26} (2009) 105015  [arXiv:0902.4634 [hep-th]].

\bibitem{Birmingham:2010mj}
  D.~Birmingham and S.~Mokhtari,
  \textit{Thermodynamic Stability of Warped $AdS_3$ Black Holes,}
  Phys.\ Lett.\ B {\bf 697} (2011) 80  [arXiv:1011.6654 [hep-th]].

\bibitem{'tHooft:1984re}
  G.~'t Hooft,
  \textit{On the Quantum Structure of a Black Hole,}
  Nucl.\ Phys.\ B {\bf 256} (1985) 727.

\bibitem{Frolov:1994zi}
  V.~P.~Frolov,
  \textit{Why the entropy of a black hole is A/4?,}
  Phys.\ Rev.\ Lett.\  {\bf 74} (1995) 3319  [gr-qc/9406037].

\bibitem{Carlip:1998wz}
  S.~Carlip,
  \textit{Black hole entropy from conformal field theory in any dimension,}
  Phys.\ Rev.\ Lett.\  {\bf 82} (1999) 2828  [hep-th/9812013].

\bibitem{Kim:2008bf}
  W.~Kim and E.~J.~Son,
  \textit{Thermodynamics of warped AdS(3) black hole in the brick wall method,}
  Phys.\ Lett.\ B {\bf 673} (2009) 90  [arXiv:0812.0876 [hep-th]].

\bibitem{York:1986it}
  J.~W.~York, Jr.,
  \textit{Black hole thermodynamics and the Euclidean Einstein action,}
  Phys.\ Rev.\ D {\bf 33} (1986) 2092.

\bibitem{Whiting:1988qr}
  B.~F.~Whiting and J.~W.~York, Jr.,
  \textit{Action Principle and Partition Function for the Gravitational Field in Black Hole Topologies,}
  Phys.\ Rev.\ Lett.\  {\bf 61} (1988) 1336.

\bibitem{Gross:1982cv}
  D.~J.~Gross, M.~J.~Perry and L.~G.~Yaffe,
  \textit{Instability of Flat Space at Finite Temperature,}
  Phys.\ Rev.\ D {\bf 25} (1982) 330.


\bibitem{York:1984wp}
  J.~W.~York, Jr.,
  \textit{Black Hole In Thermal Equilibrium With A Scalar Field: The Back Reaction,}
  Phys.\ Rev.\ D {\bf 31} (1985) 775.


\bibitem{Cai:2009ua}
  R.~-G.~Cai, L.~-M.~Cao and N.~Ohta,
  \textit{Black Holes in Gravity with Conformal Anomaly and Logarithmic Term in Black Hole Entropy,}
  JHEP {\bf 1004} (2010) 082  [arXiv:0911.4379 [hep-th]].

\bibitem{Birrell:1982ix}
  N.~D.~Birrell and P.~C.~W.~Davies,
  \textit{Quantum Fields In Curved Space},
  Cambridge, Uk: Univ.\ Pr.\ (1982) 340p

\bibitem{Deser:1993yx}
  S.~Deser and A.~Schwimmer,
  \textit{Geometric classification of conformal anomalies in arbitrary dimensions,}
  Phys.\ Lett.\ B {\bf 309} (1993) 279
  [hep-th/9302047].

\bibitem{Duff:1993wm}
  M.~J.~Duff,
  \textit{Twenty years of the Weyl anomaly,}
  Class.\ Quant.\ Grav.\  {\bf 11} (1994) 1387
  [hep-th/9308075].

\bibitem{Kazakov:1993ha}
  D.~I.~Kazakov and S.~N.~Solodukhin,
  \textit{On Quantum deformation of the Schwarzschild solution,}
  Nucl.\ Phys.\ B {\bf 429} (1994) 153
  [hep-th/9310150].

\bibitem{Kim:2012cm}
  W.~Kim and Y.~Kim,
  \textit{Phase transition of quantum corrected Schwarzschild black hole,}
  Phys.\ Lett.\ B {\bf 718} (2012) 687
  [arXiv:1207.5318 [gr-qc]].

\bibitem{Carlip:2003ne}
  S.~Carlip and S.~Vaidya,
  \textit{Phase transitions and critical behavior for charged black holes,}
  Class.\ Quant.\ Grav.\  {\bf 20} (2003) 3827
  [gr-qc/0306054].

\bibitem{Tolman:1930zza}
  R.~C.~Tolman,
  \textit{On the Weight of Heat and Thermal Equilibrium in General Relativity,}
  Phys.\ Rev.\  {\bf 35} (1930) 904.

\bibitem{Gibbons:1976ue}
  G.~W.~Gibbons, S.~W.~Hawking,
  \textit{Action Integrals and Partition Functions in Quantum Gravity,}
  Phys.\ Rev.\ D {\bf 15} (1977) 2752-2756.

\bibitem{Gibbons:1978ac}
  G.~W.~Gibbons, S.~W.~Hawking, M.~J.~Perry,
  \textit{Path Integrals and the Indefiniteness of the Gravitational Action,}
  Nucl.\ Phys.\ B {\bf 138} (1978) 141.

\bibitem{Padmanabhan:2012gx}
  T.~Padmanabhan,
  \textit{Emergent perspective of Gravity and Dark Energy,} 
  Res.\ Astron.\ Astrophys.\  {\bf 12} (2012) 891  [arXiv:1207.0505 [astro-ph.CO]].

\bibitem{Dabholkar:2012pk}
  A.~Dabholkar and S.~Nampuri,
  \textit{Lectures on Quantum Black Holes,}
  Lect.\ Notes Phys.\  {\bf 851} (2012) 165  [arXiv:1208.4814 [hep-th]].




\end{thebibliography}
\end{document}